\documentclass[10pt,twocolumn,twoside]{IEEEtran}

\usepackage{graphicx,epsfig}
\usepackage{cite}
\usepackage{stfloats}
\usepackage{amsmath}
\usepackage{amsfonts,helvet}
\usepackage{caption}
\usepackage{subcaption}
\usepackage{flushend}

\usepackage{algorithm}
\usepackage{algorithmic}

\newtheorem{theorem}{Theorem}

\newtheorem{algo}[theorem]{Algorithm}

\newtheorem{proposition}[theorem]{Proposition}

\newcounter{proposition}
\makeatletter
\newcommand{\rmnum}[1]{\romannumeral #1}
\newcommand{\Rmnum}[1]{\expandafter\@slowromancap\romannumeral #1@}
\makeatother

\begin{document}

\title{Symbol Interval Optimization for Molecular Communication with Drift}


\author{\authorblockN{Na-Rae Kim, \IEEEmembership{Student~Member,~IEEE}, Andrew W. Eckford, \IEEEmembership{Member,~IEEE}, and\\ Chan-Byoung Chae, \IEEEmembership{Senior~Member,~IEEE}}


\thanks{N. Kim and C.-B. Chae are with the School of Integrated Technology, Yonsei University, Korea. Email: \{nrkim, cbchae\}@yonsei.ac.kr.

A. W. Eckford is with the Department of Computer Science and Engineering, York University, Canada. Email: aeckford@yorku.ca.

This work was in part supported by the Ministry of Knowledge Economy under the ``IT Consilience Creative Program"  (NIPA-2014-H0201-14-1001) and the ICT R\&D program of MSIP/IITP.}}




\maketitle \setcounter{page}{1} 
%
%
%


\begin{abstract}
In this paper, we propose a symbol interval optimization algorithm in molecular communication with drift. Proper symbol intervals are important in practical communication systems since information needs to be sent as fast as possible with low error rates.
There is a trade-off, however, between symbol intervals and inter-symbol interference (ISI) from Brownian motion. Thus, we find proper symbol interval values considering the ISI inside two kinds of blood vessels, and also suggest no ISI system for strong drift models. Finally, an isomer-based molecule shift keying (IMoSK) is applied to calculate achievable data transmission rates (achievable rates, hereafter). Normalized achievable rates are also obtained and compared in one-symbol ISI and no ISI systems. 
\end{abstract}



\section{Introduction}
\label{Sec:Intro}
\begin{figure}[!t]
 \centerline{\resizebox{1\columnwidth}{!}{\includegraphics{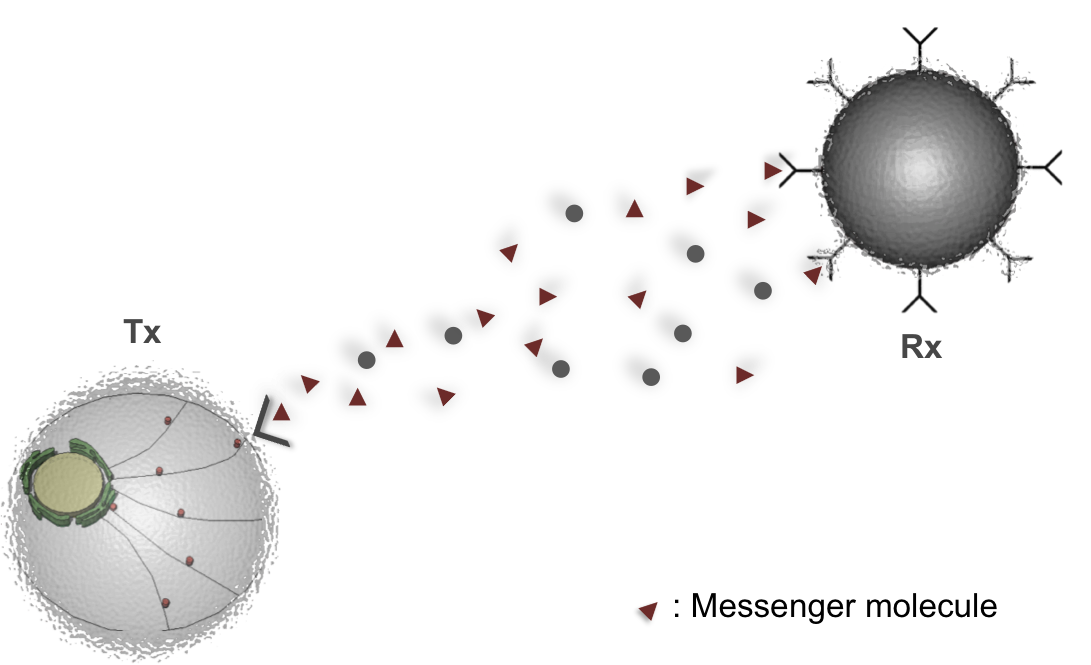}}}
  \caption{A simple case of molecular communication via diffusion with a single transmitter and a single receiver. Energy consumption model at the transmitter side is motivated by intracellular communication of eukaryotic cells as described inside the transmitter. The black circles indicate the possible noise sources present in a propagating medium. }
  \label{Fig:system}
\end{figure}

In molecular communication, information is transmitted using patterns of molecules \cite{Akyildiz_cn08,Eckford07}.  The molecules, called information/messenger molecules, can encode information using various methods: for example, in~~\cite{MOD_Kim12}, three isomer-based modulation techniques are introduced, and specific molecule sets can be chosen depending on system conditions; in other work, information is encoded in the timing or concentration of molecules~\cite{eck09,Kuran_ICC11}. 
It basically sends/receives certain information by deploying biological molecules as shown in Fig~\ref{Fig:system}. Several propagation options are possible such as diffusion-based~\cite{Max_jsac10,Akan_ncn10,Nakano_commlett12}, walkaway-based~\cite{Eckford_tnbs12,Moore_CPC06}, and flow-based~\cite{drift_Eckford12,Miorandi_NCN11} communications.

Molecular communication is a biologically-inspired solution to the problem of communication in nanoscale networks, and has been investigated in a growing body of research~\cite{Nakano_TNB2012}.
The authors in~\cite{Akyildiz_cn08, Moore_CPC06, Akyildiz_cn09} investigated several biological systems that can be background models for molecular communication systems. In addition, there has been several work to analyze maximum data transmission rates in diffusion-based molecular communication systems just as done in traditional communication systems~\cite{Kuran_ICC11,drift_Eckford12,MOD_Kim12}. Though lots of potential applications are also addressed in~\cite{Akyildiz_cn08}, normalized data rate (in bits/s) must be increased to make these applications feasible. 

Nanomedicine is a key application area for molecular communication. 
Since blood is not a stationary medium inside human body, this paper considers molecular communication in which the 
Brownian motion model incorporates drift. As can be seen in~\cite{Kuran_NCN10, Kuran_ICC11, MOD_Kim12}, Brownian motion is generally approximated as a Gaussian model. This does not hold, however, if there exists medium velocity. Thus, another model is required: the distribution of the first arrival time in Brownian motion with positive drift, i.e., medium flow direction from the transmitter to the receiver nanomachine, is given by the inverse Gaussian distribution~\cite{drift_Eckford12}.

\begin{figure*}[t]
        \centering
        \begin{subfigure}{0.4\textwidth}
                \centering
                \includegraphics[width=\textwidth]{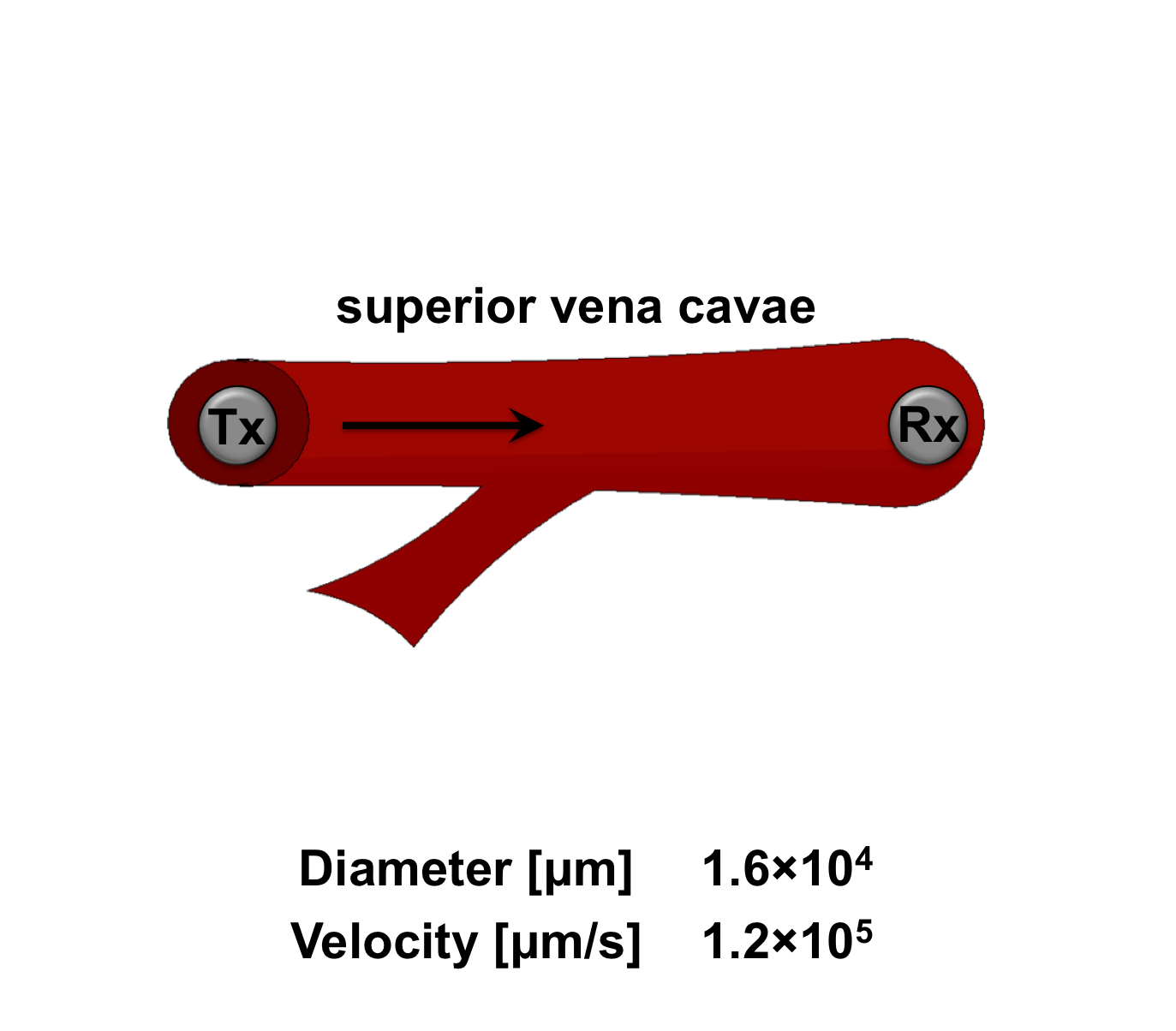}
                \caption{Inside superior vena cavae.}
                \label{Fig:cavae}
        \end{subfigure}%
        ~
        \begin{subfigure}{0.45\textwidth}
                \centering
                \includegraphics[width=\textwidth]{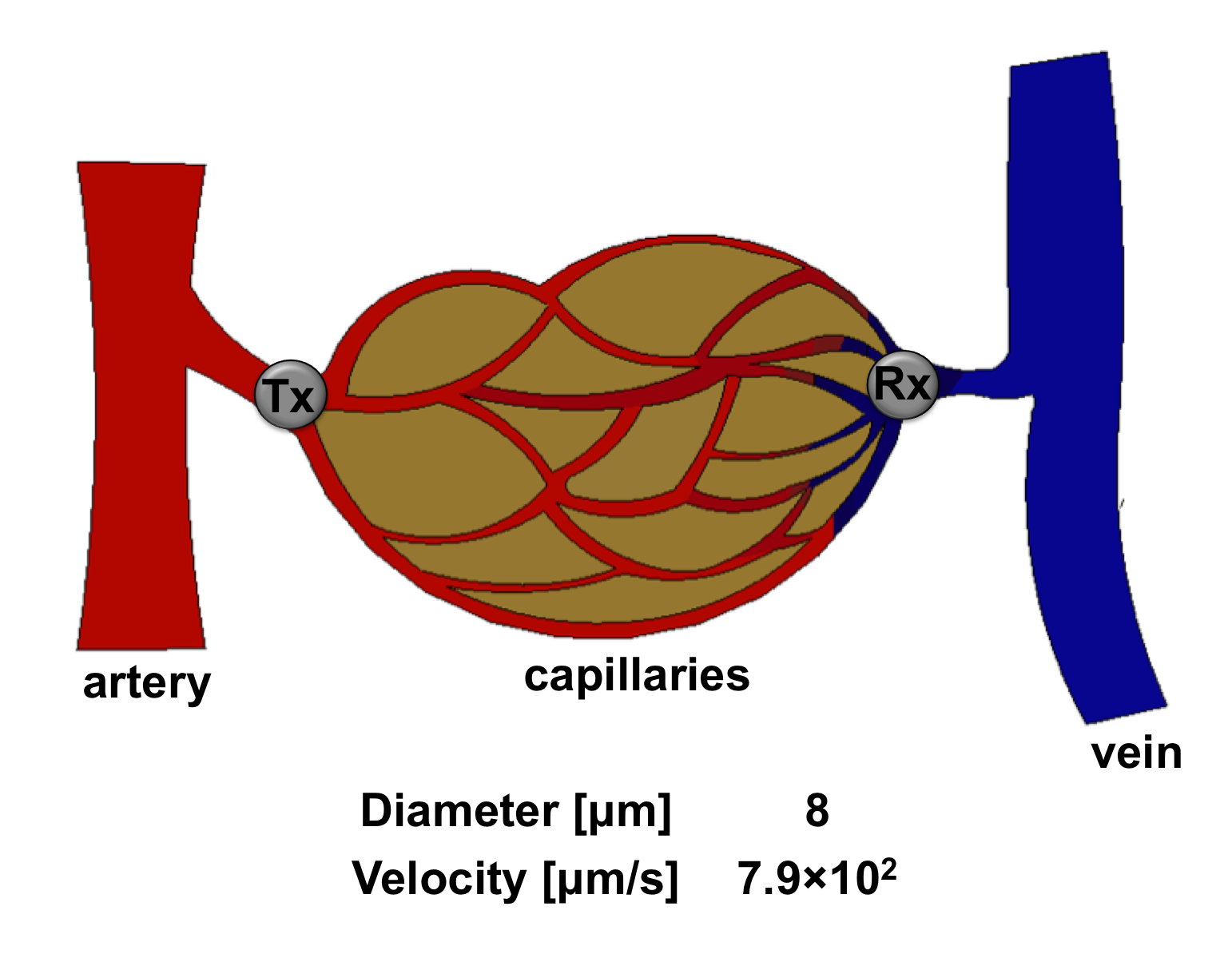}
                \caption{Inside capillaries}
                \label{Fig:cap}
        \end{subfigure}
        ~ 
        \caption{System model inside superior vena cavae and capillaries.}\label{Fig:vessels}
\end{figure*}

The analysis in \cite{drift_Eckford12} concerned the information-theoretic achievable rates, assuming an infinite symbol interval (i.e., transmission time interval), in which the receiver could wait until all transmitted molecules arrive at the receiver.
Of course, achieving the information-theoretic capacity is far beyond what is currently possible in molecular communication. It is, therefore, more realistic to focus on tasks such as symbol detection.
The main contribution of this paper is to establish the symbol interval, considering properties of the Brownian motion: to do so, it is necessary to balance the trade-off between symbol interval and ISI coming from Brownian motion (i.e., diffusion). ISI indicates the interference from undesired symbols, and molecular communication inherently has the ISI effect from the nature of diffusion~\cite{Kuran_NCN12, Pierobon_ICC12}. 
Since the hitting time pdf of molecular communication spreads in a time scale, we need an infinite symbol interval to implement ISI free systems. However, in practical systems, it is important to have a proper (or finite) symbol interval in spite of the existence of ISI. 
In this paper we
derive proper symbol intervals for molecular communication in one-symbol ISI considering both previous and current symbols. It is also possible to operate in no ISI systems when there is a strong drift in the medium. Moreover, we obtain achievable rates deploying one of the modulation techniques~\cite{MOD_Kim12}, and normalized them using the obtained symbol intervals. Our results demonstrate that transmit mode selection is also possible.

This paper is organized as follows. Section~\ref{Sec:Main} describes the system model under consideration. Section~\ref{Sec:Optm} investigates symbol interval optimizations in one-symbol ISI and no-ISI systems. We present numerical results and conclusions in Sections~\ref{Sec:Num} and~\ref{Sec:Conc}, respectively.

\section{System Model}
\label{Sec:Main} 
We assume a point-to-point communication inside a human body, i.e., there exists a single transmitter and a single receiver inside one of blood vessels.
The transmitter nanomachine transmits a specific number of messenger molecules at the beginning of a symbol, and waits for the next release of molecules until the ISI becomes insignificant (until previously sent molecules fade away enough). That is, we could wait for all the molecules to be received at the receiver side, or set a proper threshold value of which percentage of molecules has to be waited to ignore the ISI effect. 

The molecules are affected by both diffusion and blood velocity inside which they travel (e.g., inside capillaries or cavae). 
We will assume that the symbol interval is determined by drift velocity and distance between the communication terminals (i.e., transmitter and receiver).

For the Brownian motion with medium velocity, we apply the inverse Gaussian model as shown below~\cite{drift_Eckford12,Kim_JSAC2014,Birkan_CL2014}: 
\begin{align}
f(t;\mu, \lambda) = \Big({\frac{\lambda}{2 \pi t^3}\Big)}^{1/2} \exp\Big(\frac{-\lambda{(t-\mu)}^2} {2 {\mu}^2 t} \Big)
\label{IG}
\end{align}
where, $f(t;\mu, \lambda)$ represents a probability density function (PDF) of the first hitting time $t$ with a mean of $\mu$ and a shape parameter of $\lambda$. 
It can be well approximated by a Gaussian with increasing value of $\lambda$. 
 In~\cite{drift_Eckford12}, they derived a pdf of hitting time (i.e., the time when the molecules first hit the receiver nanomachine, or absorption time) as follows: 
\begin{align}
f(t) = \frac{d}{\sqrt{4 \pi D t^3}} \exp \Big(\frac{-{(vt-d)}^2}{4Dt}\Big)
\label{IG_Eckford}
\end{align}
where, $d$ represents a distance between a transmitter and a receiver, $D$ represents a diffusion coefficient of a Brownian particle, and $v$ indicates medium velocity. By comparison of (\ref{IG}) and (\ref{IG_Eckford}), we can represent the mean and shape parameter in terms of the distance, diffusion coefficient, and the medium velocity, i.e.,
\begin{align}\begin{split}
&\mu = \frac{d}{v}~\text{(mean)}, \\
&\lambda = \frac{d^2}{2D}~\text{(shape parameter)}.
\label{IG_coeff}
\end{split}\end{align}

From the basic property of the inverse Gaussian distribution and (\ref{IG_coeff}), a cumulative density function (CDF) for the hitting time $t$ can be obtained as 
\begin{align}\begin{split}
&F(t) = \Phi\Bigg( \sqrt{\frac{\lambda}{t}} \Big(\frac{t}{\mu}-1\Big) \Bigg)+\exp\Big(\frac{2\lambda}{\mu}\Big) \Phi\Bigg(-\sqrt{\frac{\lambda}{t}} \Big(\frac{t}{\mu}+1\Big) \Bigg)\\
&= \Phi\Bigg( \frac{d}{\sqrt{2Dt}} \Big(\frac{vt}{d}-1\Big) \Bigg)+\exp\Big(\frac{vd}{D}\Big) \Phi\Bigg(-\frac{d}{\sqrt{2Dt}} \Big(\frac{vt}{d}+1\Big) \Bigg)
\end{split}\nonumber\end{align}
where $F(t)$ represents a cdf of hitting time and $\Phi(\cdot)$ represents a standard Gaussian cdf. That is, $F(t)$ means the probability of hitting the receiver located at a fixed point within time $t$. Thus, we can define this as a hitting probability, $P_\text{hit}$, so we have $F(t) = P_\text{hit}$. Since the hitting probability is mainly a function of medium velocity, time and distance between a transmitter and a receiver, we use $P_\text{hit}$ for $P_\text{hit}(v,d,t)$, hereafter, assuming a constant diffusion coefficient for a specific messenger molecule.

We apply the similar system models that are used in~\cite{MOD_Kim12} to analyze in practical systems. Thus, it is assumed that the number of messenger molecules that hit the receiver follows a binomial distribution ($n$ number of trials with a probability of $P_\text{hit}$ for each molecule), and is approximated as a normal distribution as shown below. Here, we can safely approximate the binomial distribution as the normal distribution when $n$ is large enough, and $P_{hit}$ is not close to 1 or 0.
\begin{align}\begin{split}
&\mathrm{Binomial} (n, P_\text{hit})\sim \mathcal{N}(nP_\text{hit}, nP_\text{hit}(1-P_\text{hit})).
 \end{split}
\label{normal}  \end{align}
Also, we use the same energy model motivated by intra-cellular communication of eukaryotic cells. In short, at the transmitter side, molecules synthesized by nucleus are encapsulated inside vesicles, which travel through microtubules. Next, vesicles carry them to the boundary of the transmitter, and release the molecules into the medium. The processes are simplified in Fig.~\ref{Fig:system} at the transmitter side, and the energy consumption in each step is calculated from biological references as specified in~\cite{MOD_Kim12}.

\begin{figure}[!t]
 \centerline{\resizebox{0.95\columnwidth}{!}{\includegraphics{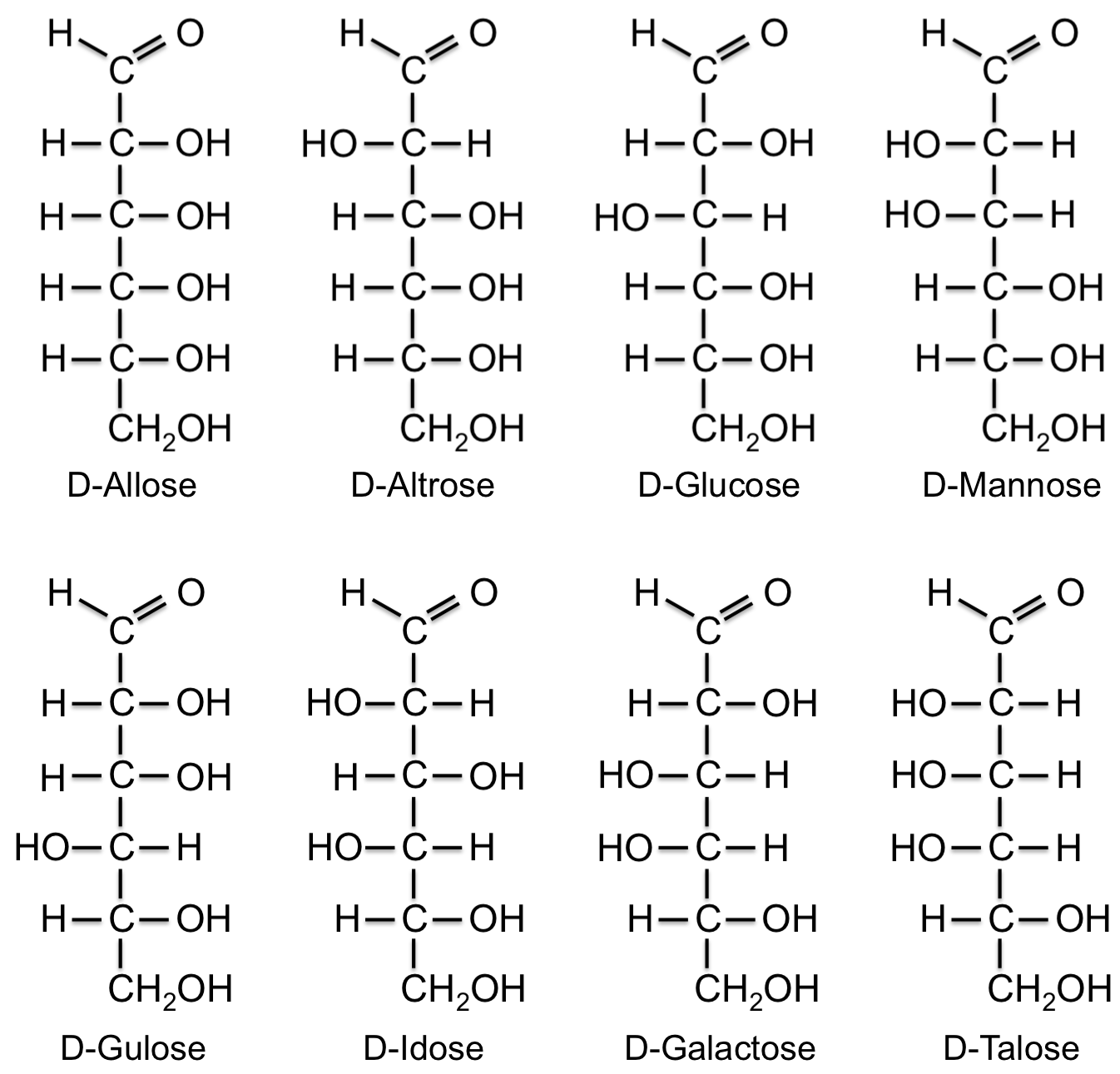}}}
  \caption{Some examples of molecule sets that can be used for Q-IMoSK: different isomers of hexoses.}
  \label{Fig:hexoses}
\end{figure}

\begin{figure}[!t]
 \centerline{\resizebox{1\columnwidth}{!}{\includegraphics{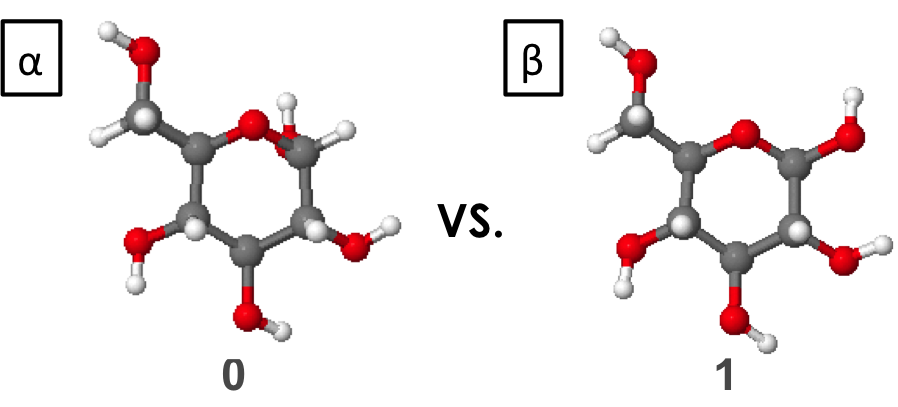}}}
  \caption{B-IMoSK system using two different isomers of hexoses ($\alpha$ and $\beta$ type), and each molecule represents binary symbols, `0' and `1'. Jmol is used to plot the figures~\cite{Jmol}.}
  \label{Fig:IMoSK}
\end{figure}

Lastly, the IMoSK is applied to calculate the achievable rates among other modulation techniques. In our prior work~\cite{MOD_Kim12}, we proposed several modulation methods such as ICSK (isomer-based concentration shift keying), IRSK (isomer-based ratio shift keying), and IMoSK. Each one has its own advantages, and we choose one depending on system conditions. Here, we use the IMoSK that represents different information using different kinds of molecules (i.e., isomers, molecules composed of the same number and types of atoms). By utilizing our proposed isomers (e.g., hexoses), we obtain several advantages such as safety, convenience, low complexity, and systematic analysis.
Some examples (isomers of hexoses) are described in Fig.~\ref{Fig:hexoses}, and they exist as a ring type in aqueous solution as shown in Fig.~\ref{Fig:IMoSK}.
In Fig.~\ref{Fig:IMoSK}, it shows the simplest case of IMoSK systems, binary-IMoSK (B-IMoSK), and each molecule can represent binary symbols, i.e., `0' and `1'. To deploy quadrature-IMoSK (Q-IMoSK), four different isomers are required representing four different symbols.

\section{Symbol Interval Optimization}
\label{Sec:Optm}
\begin{figure}[!t]
 \centerline{\resizebox{1.1\columnwidth}{!}{\includegraphics{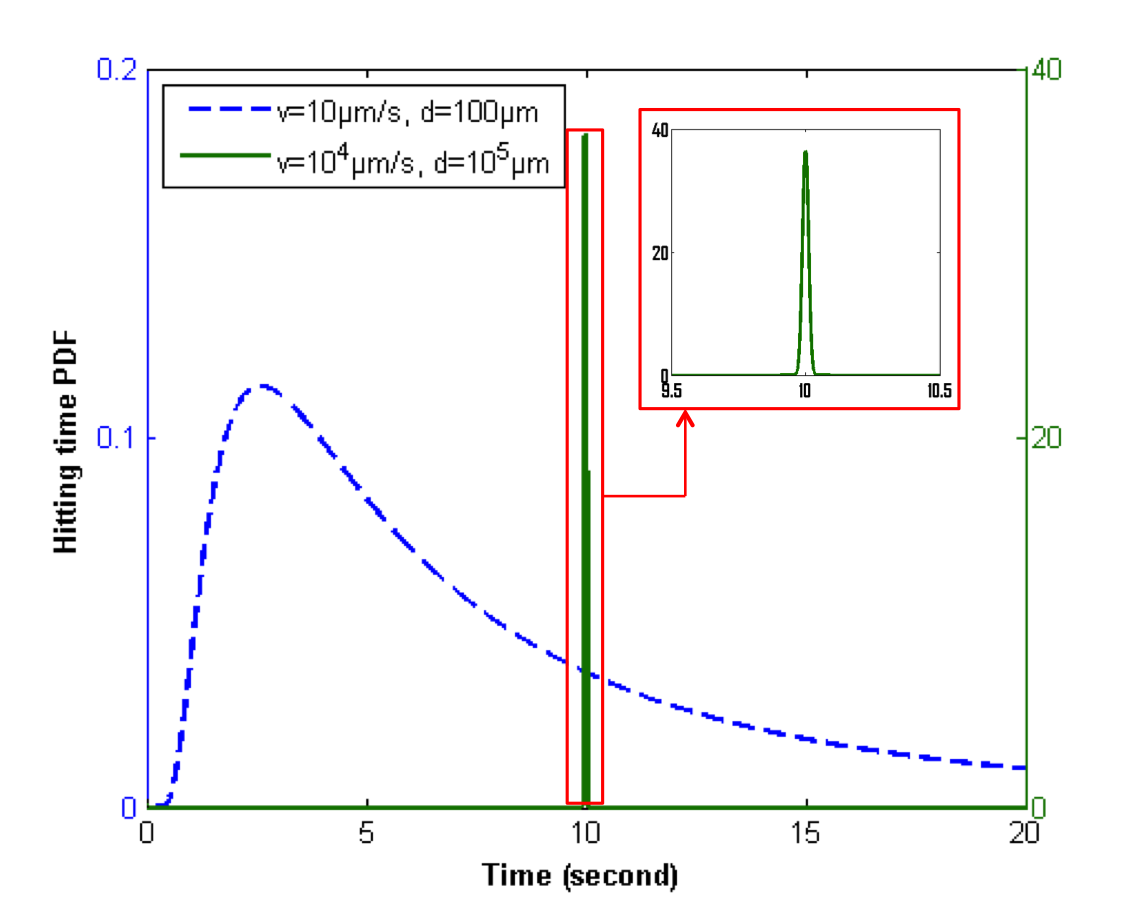}}}
  \caption{Comparison of hitting time pdf for low ($v=10 \mu m/sec$, $d=100\mu m$) and high ($v=10^4 \mu m/sec$, $d=10^5 \mu m$) velocity models.}
  \label{Fig:IG}
\end{figure}

As explained in Section~\ref{Sec:Intro}, practical communication systems require a proper symbol interval to reliably transmit data. Brownian motion, however, inherently has ISI, and it affects to determine symbol intervals. The system also assumes the first passage process, which means that the messenger molecules disappear after decoding. Therefore, decoded molecules at the present symbol interval do not affect the next symbol, and do not generate ISI.
In Fig.~\ref{Fig:IG}, the inverse Gaussian distribution is shown in two cases as examples: $\rmnum{1}$) when the velocity is $10\mu m/sec$ and the distance is $100\mu m$, $\rmnum{2}$) when the velocity is $10^4\mu m/sec$ and the distance is $10^5\mu m$. 

The first pdf in Fig.~\ref{Fig:IG} has a long tail since diffusion effect dominates with small velocity. In this case, the symbol interval goes infinite to ignore ISI that means it is almost impossible to wait for all the transmitted molecules. Therefore, we have to tolerate ISI to some level meaning the previous symbol affects the current symbol, and it will be discussed in the following (Section~\ref{SubSec:ISI}) that how much we tolerate.
The second pdf in Fig.~\ref{Fig:IG} shows an almost impulse peak due to the higher drift velocity, and it makes no ISI system feasible. It will be specified in Section~\ref{SubSec:noISI}.

\begin{figure*}[!t]
 \centerline{\resizebox{1.9\columnwidth}{!}{\includegraphics{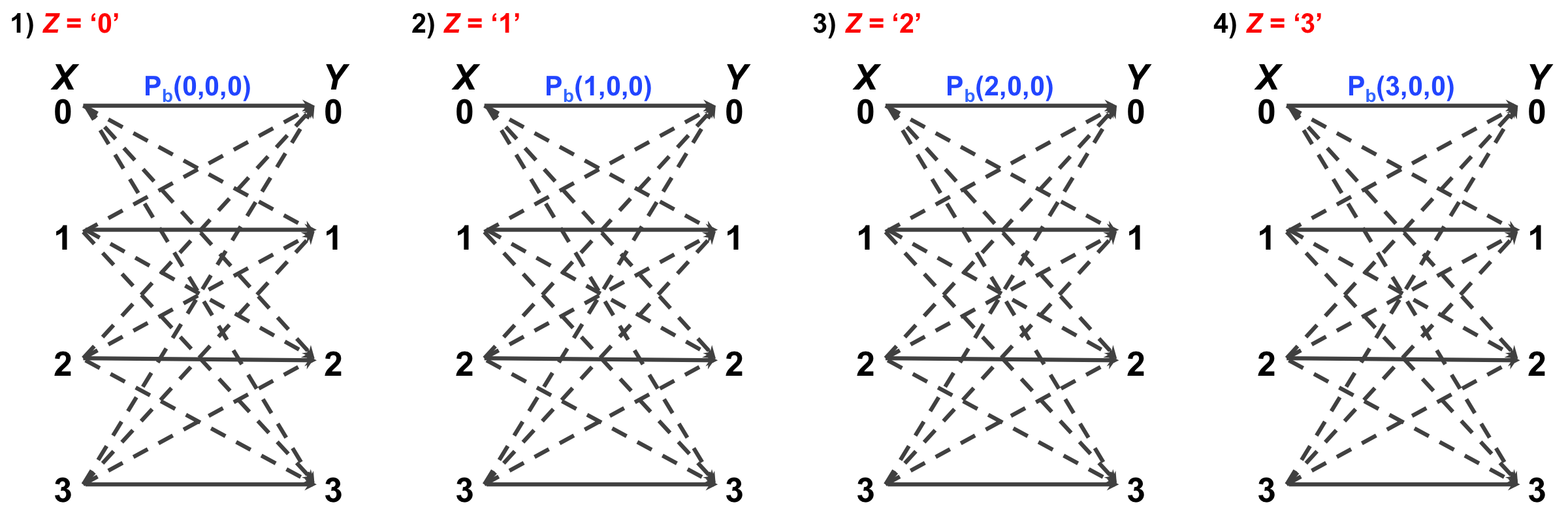}}}
  \caption{The number of possible cases in Q-IMoSK system. There exists 64 (4$\times$4$\times$4) different cases to be considered.}
  \label{Fig:Qchannel}
\end{figure*}

\subsection{One-symbol ISI system}  
\label{SubSec:ISI}
Generally, the inverse Gaussian distribution has a long tail, which means it is impractical to wait for all the messenger molecules to hit the receiver:
some molecules will arrive in intervals later than the one
in which they were transmitted, leading to ISI. 
We first have to define up to which percentage level ISI can be ignored. Here, motivated by~\cite{Kuran_NCN10}, it is assumed that ISI effect becomes insignificant if (\ref{ISI}) is satisfied .
\begin{align}\begin{split}
&1.~ F(T_s) > A,\\
&2.~F(3 T_s)-F(2 T_s) = P_{\text{hit}}(3 T_s) - P_{\text{hit}}(2 T_s) \\
&< \epsilon\times F(T_s)= \epsilon\times P_{\text{hit}}(T_s),
\label{ISI}
\end{split}\end{align}
where $A$ is large and $\epsilon$ is small enough. It indicates that hitting probabilities at $T_s$ is high enough and do not significantly increase after $2T_s$. We can control $A$ and $\epsilon$ depending on system conditions (e.g., medium velocity and distance), and here, from test simulations, they are set to be 0.8 and 0.001, respectively. Again, the hitting probability for the current symbol is greater than 0.8, and the increase of the received molecules from $2T_s$ to $3T_s$ is smaller than 0.001 times of the hitting probability. 
Thus, $T_s$ value satisfying the condition becomes the desired symbol interval, ${T_s}^*$, and the cdf of ${T_s}^*$ is the percentage of molecules that have to be waited for before the transmitter starts the next release.

Now, we apply the obtained symbol interval to derive normalized achievable rates of the system. As specified in~\cite{MOD_Kim12}, mutual information between transmitted and received symbols ($X$ and $Y$, respectively) can be calculated using probabilities of $X$ sent and $Y$ received, $P_a(X,Y)$. It can be represented as the sum of $P_b(Z,X,Y)$, the probability of $X$ sent, $Y$ received, and $Z$ previously sent. Since IMoSK is used for our modulation method, symbols, $X$, $Y$, and $Z$, can be differentiated by deploying different messenger molecules, or isomers. 
Therefore, the achievable rates of IMoSK system can be obtained as follows:
\begin{align}\begin{split}
\text{If $X$=$Y$}\text{ (i.e.,}&\text{ if $X$ sent, $X$ received.)},\\
P_a(X,Y)&=\sum_{Z}P_b(Z,X,Y)\\
&={\Big(\frac{1}{4}\Big)}^2\Bigg[P(N_p+N_c+N_n\geq\tau)P(N_n<\tau)\\
&+3\Big(P(N_c+N_n\geq\tau)P(N_p+N_n<\tau)\Big)\Bigg].
\nonumber
\end{split}
\end{align}
\begin{align}\begin{split}
\text{If $X$}{\neq}\text{$Y$} \text{ (i.e.,}&\text{ if $X$ sent, $Y$ received)},\\
P_a(X,Y)&=\sum_{Z}P_b(Z,X,Y)\\
&={\Big(\frac{1}{4}\Big)}^2\Bigg[P(N_p+N_n\geq\tau)P(N_c+N_n<\tau)\\
&+3\Big(P(N_n\geq\tau)P(N_p+N_c+N_n<\tau)\Big)\Bigg].
\end{split}
\nonumber\end{align}
Here, $\tau$ indicates threshold, a minimum requirement of the number of molecules to be detected at the receiver side. The $N_p$ denotes the number of molecules previously transmitted, but currently received. Also, $N_c$ and $N_n$ are the number of currently transmitted molecules and noise molecules, respectively.

As mentioned in Section.~\ref{Sec:Main}, Q-IMoSK is applied for modulation technique described in Fig.~\ref{Fig:Qchannel}. Therefore, there exist four possible cases (i.e., different types of molecules) for each symbol, $X$, $Y$, and $Z$, and 64 cases are to be considered in total. 
$P_b(Z,X,Y)$ for each $Z$ value can be summed up to calculate $P_a(X,Y)$, and the $N_p$ and $N_c$ can be approximated as a normal distribution by (\ref{normal}). These also result in the form of $Q$ function (i.e., the tail probability of the normal distribution) using the relationship between $Q$ function and the normal distribution.
\footnote{$Q$ function is defined as $Q(x)=1/\sqrt{2\pi}\int_x^{\infty} \exp(-u^2/{2})du$ or $Q(x)=\frac{1}{2}\text{erfc}({{x}/{\sqrt{2}}})$.}

\begin{figure}[!t]
 \centerline{\resizebox{1.09\columnwidth}{!}{\includegraphics{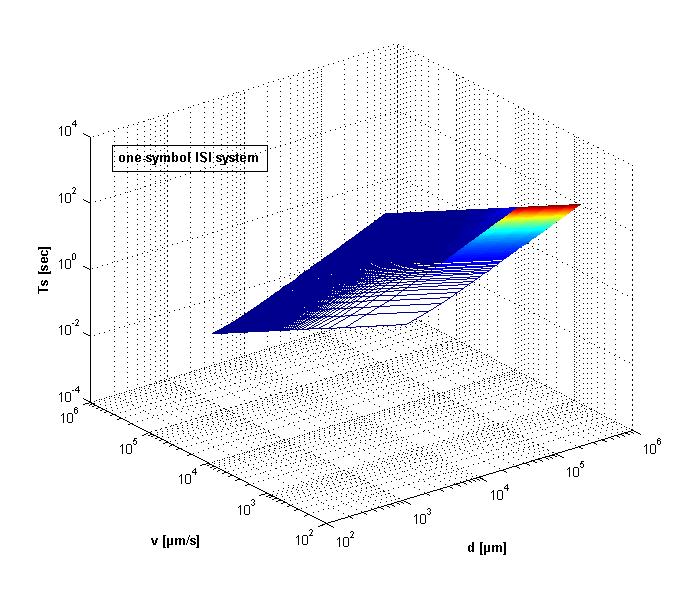}}}
  \caption{Desired symbol interval in terms of both velocity and distance in one symbol ISI system. It is shown in logarithmic scale.}
  \label{Fig:ts_ISI}
\end{figure}

\begin{table*}[!b]
\caption{Desired symbol interval ($T_s$, $sec$) in one symbol ISI  and no ISI systems inside capillaries and superior vena cavae.}
\begin{center}
\begin{tabular}{|c|c|c|c|c|}
\hline
& Velocity ($\mu m/s$) & Distance ($\mu m$) & $T_s$ for one symbol ISI & $T_s$ for no ISI  \\ \hline \hline
Capillaries (Fig.~\ref{Fig:cap}) & 7.9$\times10^2$ &7.9$\times10^2$ & 2.0640 & 4.1110 \\ \hline
Superior vena cavae (Fig.~\ref{Fig:cavae}) & 1.2$\times10^5$ & 1.2$\times10^5$ & 1.0390 & 1.1490\\ \hline
No drift&0 & 16 & 5.9 & $\infty$\\ \hline
\end{tabular}
\end{center}
\label{symbol_interval}
\end{table*}

\subsection{No ISI system}
\label{SubSec:noISI}
In Brownian motion with strong drift, the hitting time pdf is much sharper than the weak drift model as shown in Fig.~\ref{Fig:IG}, which means it is expected to be practically possible to wait for all the molecules transmitted to be received. In that case, the probabilities of $X$ sent and $Y$ received, $P_a(X,Y)$, are calculated in a different way with Section~\ref{SubSec:ISI}. In other words, we no longer consider the previous symbol since the current symbol is sent only after all the previously sent molecules arrive at the receiver side. $P_a(X,Y)$ values for a Q-IMoSK system with no ISI can be calculated as follows:
\begin{align}
\text{If }X=Y,~P_a(X,Y)=\frac{1}{4} P(N_c+N_n \geq \tau)P(N_n<\tau).\nonumber\\
\text{If }X\neq Y,~P_a(X,Y)=\frac{1}{4} P(N_n \geq \tau)P(N_c+N_n<\tau).\nonumber
\end{align}

Symbol interval values are expected to become higher compared to the one-symbol ISI system since all the transmitted molecules have to be waited for. The error probabilities, however, are expected to be lower with no ISI. Therefore, the two systems are in a trade-off relationship, and we can select a proper system depending on system conditions (e.g., medium velocity or transmit power).

\subsection{Transmit mode selection}
We assume that the transmitter inside blood vessels is equipped with a specific sensor that measures the environment (i.e., medium conditions), and also knows the distance to the receiver as pre-information. In that case, the transmitter has all the information (medium velocity, distance, and signal power) to calculate proper symbol intervals, and can determine in which transmit mode it has to be operated. For instance, with strong medium drift, no ISI system will be more appropriate especially with low SNR, i.e., low signal power. Note that the receiver does not need to know this symbol interval. This will be numerically proven in Section~\ref{Sec:Num}.

\section{Numerical Results}
\label{Sec:Num}
Hexoses are used for simulations in this work. As explained in~\cite{MOD_Kim12}, isomers are one of the best candidates to be messenger molecules for several reasons, and hexoses are a good example for them. These have 32 different isomers that can be deployed in the selected modulation technique, and we can do more systematic analysis.
Diffusion coefficient can be calculated from $ D = \frac{K_b T}{6\pi \eta r_{mm}} \label{D}$ where $r_{mm}$ represents the radius of messenger molecules, $K_b$ is Boltzmann constant, $T$ is temperature, and $\eta$ means viscosity. 
Desired symbol interval values are obtained inside specific blood vessels according to the algorithm suggested in Section~\ref{Sec:Optm}, and the normalized achievable rates are calculated. 
Here, we define the achievable rate $R$ that maximizes the mutual information $I(X;Y)$ as follows:
\begin{align}\begin{split}
&I(X;Y)=\sum_X \sum_Y P_a(X,Y)\log_2{\frac{P_a(X,Y)}{P(X)P(Y)}}, \\
&R =\max_{\tau} I(X;Y)  
\end{split} \label{mi}
\end{align}
where, $P(X)$ and $P(Y)$ are the probabilities of events $X$ (the transmitted symbol) and $Y$ (the received symbol).

\begin{table}[!t]
\caption{Simulation parameters.}
\begin{center}
\begin{tabular}{|c|c|}
\hline
Parameters & Values \\
\hline\hline
Radius of the hexoses~\cite{hexose_radi} & 0.38 $nm$ \\ \hline
Temperature (body temperature) & 36.5 $^{\circ}C$ = 310$K$ \\ \hline
Viscosity of blood at body temperature~\cite{vis} & 2.46$\times$10$^{-3}$$kg/sec \cdot m$\\ \hline
$D$ of hexoses in blood &242.78 ${\mu m}^2/sec$ \\ \hline
\end{tabular}
\end{center}
\label{parameters}
\end{table}

Table~\ref{symbol_interval} shows the desired symbol intervals  inside capillaries and superior vena cavae compared with those from no drift model. Blood has certain velocities inside vessels, and the different symbol intervals should be deployed with different velocities. As velocity increases, $T_s$ decreases since the molecules arrive at the receiver side in a shorter time duration. Also in cavae, there is less difference between one-symbol and no ISI system since the velocity is much higher. Moreover, as can be easily expected, no drift model has much higher $T_s$ resulting in much lower normalized achievable rates compared with drift models.
Fig.~\ref{Fig:ts_ISI} represents the desired symbol interval as a function of medium velocity and distance in one-symbol ISI systems in logarithmic scale. As shown in the figures, $T_s$ generally increases with distance and decreases with velocity, and obviously it has larger values in no ISI system (not shown in this paper to avoid duplication).

\begin{figure}[!t]
 \centerline{\resizebox{1.13\columnwidth}{!}{\includegraphics{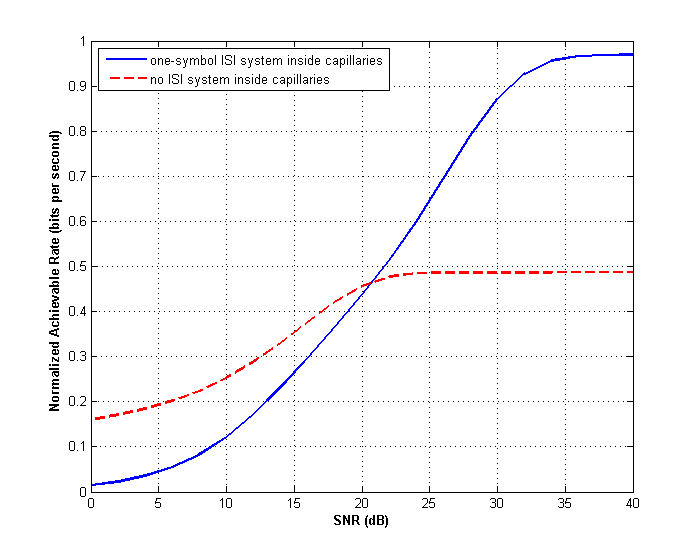}}}
  \caption{Achievable rate comparison of one symbol ISI and no ISI system inside capillaries. The rate is normalized to the desired symbol interval. System parameters are given in Table~\ref{symbol_interval}.}
  \label{Fig:rate_cap}
\end{figure}

\begin{figure}[!t]
 \centerline{\resizebox{1.13\columnwidth}{!}{\includegraphics{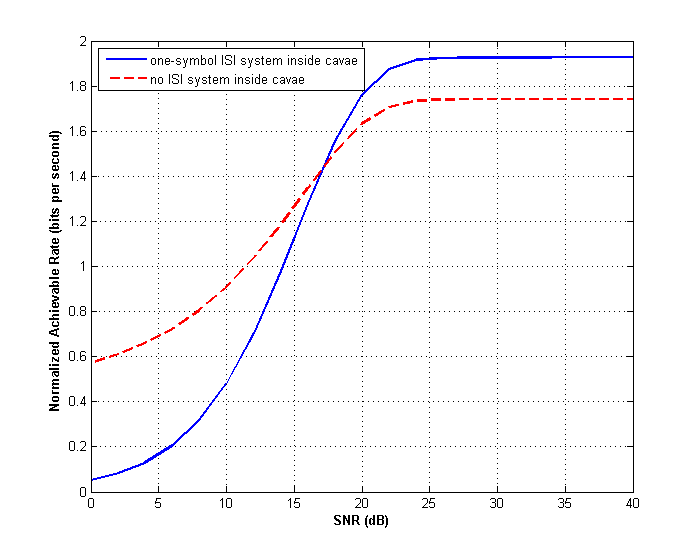}}}
  \caption{Achievable rate comparison of one symbol ISI and no ISI system inside superior vena cavae. The rate is normalized to the desired symbol interval value. System parameters are given in Table~\ref{symbol_interval}.}
  \label{Fig:rate_cavae}
\end{figure}

\begin{figure}[h]
 \centerline{\resizebox{1.13\columnwidth}{!}{\includegraphics{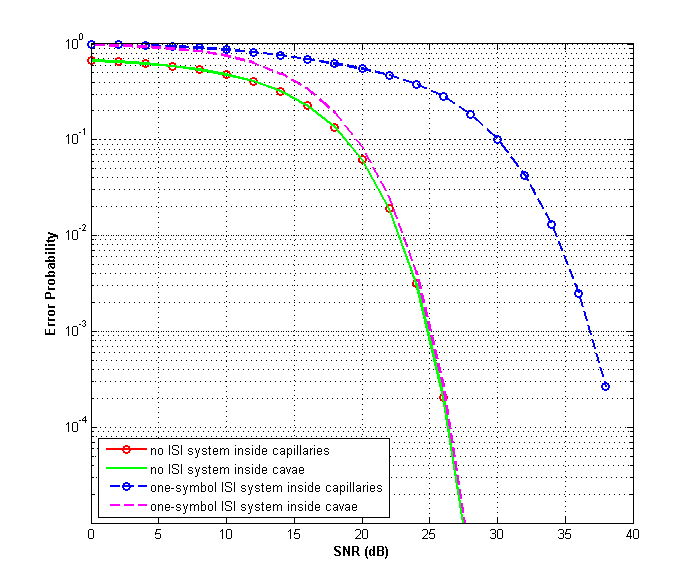}}}
 \caption{Error probability comparisons of the four different systems with differenct transmit mode and drift velocities.}
 \label{Fig:BER}
\end{figure}

Lastly, Figs.~\ref{Fig:rate_cap} and \ref{Fig:rate_cavae} compare the normalized achievable rates in one symbol ISI and no ISI systems inside capillaries and superior vena cavae, respectively. Also, Fig.~\ref{Fig:BER} compares the error probability performances of the four different cases. Here, SNR is defined as the signal to noise power ratio, and the signal power is calculated using the number of transmitted molecules (to be converted into a power unit by energy model) and the hitting probability. Also, we assumed a random noise, and made some variations to noise power with fixed signal power to adjust SNR values.

In Figs.~\ref{Fig:rate_cap} and \ref{Fig:rate_cavae}, there are cross points between the two systems' performances. No ISI system shows a better performance in lower SNR region, and one symbol ISI system is better in higher SNR region. Thus, for a molecular communication with drift, a transmitter can select one of the systems with different SNR depending on velocity and distance values as explained in Section~\ref{Sec:Optm}. That means we can achieve the envelope of the two lines in the figure through symbol interval optimization. 
Moreover, inside cavae, the difference between the two systems in higher SNR region is much smaller than inside capillaries. That means even for higher SNR, no ISI system inside cavae shows a good performance due to the stronger drift. 

In Fig.~\ref{Fig:BER}, the error probabilities of the four different systems are calculated using the normalized bits per symbol or bits per second. As expected, one-symbol ISI system inside capillaries represent the highest error probability through the all SNR region since capillaries do not have the strong enough velocity inside. One-symbol ISI system inside cavae converges to the two no ISI systems as SNR increases. Also, no ISI systems both inside cavae and capillaries show very little difference because the two systems select the long enough symbol intervals to mitigate ISI.

%


\section{Conclusions}
\label{Sec:Conc}
The optimal symbol interval has been an important open problem in molecular communication, complicated by the presence of ISI. In this paper, we proposed an symbol interval optimization method in one-symbol ISI and no ISI systems considering a drift velocity, which have not been addressed before. 
In addition, the desired symbol intervals were applied into the Q-IMoSK system to obtain normalized achievable rates. Thus, we can determine symbol interval values in which the system can be optimized in terms of ISI. We also suggested a transmission mode selection depending on SNR when there is a cross point between one-symbol and no ISI systems. 
For future work, we will prove that~(\ref{ISI}) is reasonable both in mathematical and experimental ways. It is also possible to expand the system with multiple input (transmitters) and multiple output (receivers), i.e., molecular MIMO. Some advanced ISI mitigation techniques for molecular communication (e.g., molecular OFDM) will also be investigated.


\bibliographystyle{IEEEbib}

\bibliography{references_symbol_interval}

\begin{IEEEbiography}{Na-Rae Kim}
(S'12) received her B.S. degree in Chemical Engineering from Yonsei University, Korea in 2011. She is now with the School of Integrated Technology at the same university and is working toward the Ph.D. degree. She was an exchange student at University of California, Irvine in USA in 2009. 

Ms. Kim was the recipient of the travel grant from the IEEE International Conference on Communications in 2012, 2014 and the Gold Prize (1st) in the 19th Humantech Paper Award. 
\end{IEEEbiography}

\begin{IEEEbiography}{Andrew Eckford}
(M'96 - S'97 - M'04) is originally from Edmonton,
AB, Canada. He received the B.Eng. degree in
electrical engineering from the Royal Military
College of Canada, Kingston, ON, Canada, in 1996,
and the M.A.Sc. and Ph.D. degrees in electrical
engineering from the University of Toronto, Toronto,
ON, Canada, in 1999 and 2004, respectively.
He is currently an Associate Professor of Computer Science and Engineering at York University,
Toronto.
Dr. Eckford is also the Chair of the IEEE ComSoc
Emerging Technical Subcommittee on Nanoscale, Molecular, and Quantum
Networking.
\end{IEEEbiography}

\begin{IEEEbiography}{Chan-Byoung Chae}
(S'06 - M'09 - SM'12) is an Assistant Professor in the School of Integrated Technology, College of Engineering, Yonsei University, Korea. He was a Member of Technical Staff (Research Scientist) at Bell Laboratories, Alcatel-Lucent, Murray Hill, NJ, USA from 2009 to 2011. Before joining Bell Laboratories, he was with the School of Engineering and Applied Sciences at Harvard University, Cambridge, MA, USA as a Post-Doctoral Research Fellow. He received the Ph. D. degree in Electrical and Computer Engineering from The University of Texas (UT), Austin, TX, USA in 2008, where he was a member of the Wireless Networking and Communications Group (WNCG).

Prior to joining UT, he was a Research Engineer at the Telecommunications R\&D Center, Samsung Electronics, Suwon, Korea, from 2001 to 2005. He was a Visiting Scholar at the WING Lab, Aalborg University, Denmark in 2004 and at University of Minnesota, MN, USA in August 2007. While having worked at Samsung, he participated in the IEEE 802.16e standardization, where he made several contributions and filed a number of related patents from 2004 to 2005. His current research interests include capacity analysis and interference management in energy-efficient wireless mobile networks and nano (molecular) communications. He serves as an Editor for the \textsc{IEEE Trans. on Wireless Communications}, \textsc{IEEE Trans. on Smart Grid}, and \textsc{Jour. of Comm. Networks}. He is also an Area Editor for the \textsc{IEEE Jour. Selected Areas in Communications} (nano scale and molecular networking). He is an IEEE Senior Member.

Dr. Chae was the recipient/co-recipient of the IEIE/IEEE Joint Award for Young IT Engineer of the Year in 2014, the Haedong Young Scholar Award, the IEEE Signal Processing Magazine Best Paper Award in 2013, the IEEE ComSoc AP Outstanding Young Researcher Award in 2012, the IEEE Dan. E. Noble Fellowship Award in 2008, the Gold Prize (1st) in the 14th/19th Humantech Paper Award, and the KSEA-KUSCO scholarship in 2007. He also received the Korea Government Fellowship (KOSEF) during his Ph.D. studies.
\end{IEEEbiography}

\end{document}